\documentclass[10pt,prl,aps,twocolumn,showpacs,floatfix]{revtex4}
\usepackage{graphicx}
\usepackage{amsmath}
\usepackage{bm}
\usepackage{color}

\begin{document}
\draft
\title{Asymptotic Freedom in Quantum Magnets}
\title{ Asymptotic Freedom in Quantum 
Antiferromagnet TlCuCl$_3$.}
\author{H. D. Scammell and O. P. Sushkov }
\affiliation{School of Physics, The University of New South Wales,
  Sydney, NSW 2052, Australia}
\date{\today}
\begin{abstract}
Phase transitions in isotropic quantum antiferromagnets are described by an O(3) nonlinear quantum field theory. In three dimensions, the fundamental property of this theory is logarithmic scaling of the coupling constant. At the quantum critical point the coupling asymptotically vanishes and the quasiparticles become free. This logarithmic decay of the coupling constant has never been observed. In this paper, we derive finite temperature properties of the field theory and use our results to analyze the existing data on the real antiferromagnet TlCuCl$_3$. Including finite temperatures into the theory, we find that agreement between theory and experiment is sufficiently sensitive to unambiguously identify the asymptotic decay of the coupling constant. We also comment on the unique possibility to study Landau pole-physics in quantum magnets.
\end{abstract}

\maketitle

Asymptotic freedom  plays a crucial role in Quantum Chromodynamics.
The freedom, which is due to non-abelian gauge fields,
means a logarithmic decay of the coupling constant at high energies.
Ultimately at infinite energies particles do not 
interact, this is ultraviolet
asymptotic freedom \cite{Gross,Politzer}.
In three dimensional (3D) non-gauge quantum field theories
as well as in abelian gauge theories, the coupling constant
decays logarithmically at low energies \cite{Landau}. However, usually 
this decay is terminated because of a low energy cutoff. For example 
in Quantum Electrodynamics the cutoff is due to the rest energy of the electron.
The low energy logarithmic decay can be observed only at a quantum critical
point (QCP) where the cutoff energy is zero. It is known to the theory of quantum magnetism that certain antiferromagnetic systems can be driven to a QCP, and that at such a point the characteristic low energy magnetic excitations become non-interacting. At the QCP they experience ``infrared asymptotic freedom''. Quantum antiferromagnets in the vicinity of a QCP therefore provide a perfect testing ground for such behaviour.

The low energy logarithmic decay of the coupling constant at a QCP can, in principle, be observed at zero temperature. 
The zero temperature case is well understood theoretically, a 
3D {\it quantum} antiferromagnet at zero temperature 
is equivalent to a 4D {\it classical} antiferromagnet at finite temperature.
The 3D QCP corresponds to the N\'eel transition in the 4D case.
Thermodynamic quantities scale as powers of the running 
coupling constant \cite{Zinn-Justin}.
Unfortunately, the existing zero temperature experimental data are not sufficient 
to pin down the logarithmic scaling. On the other hand combined zero and nonzero
temperature data on TlCuCl$_3$ 
 \cite{Merchant, Ruegg2004, Ruegg2008}
provide an excellent opportunity to search for fingerprints of asymptotic freedom at the QCP.
To perform this search we develop a theory of the QCP which accounts for both
quantum and thermal fluctuations. After that we compare the theoretical predictions
with experimental data. The comparison unambiguously indicates a
logarithmically running coupling constant.

%%%%%%%%%%%%%%%%
\begin{figure}[t!]
 {\includegraphics[width=0.45\textwidth,clip]{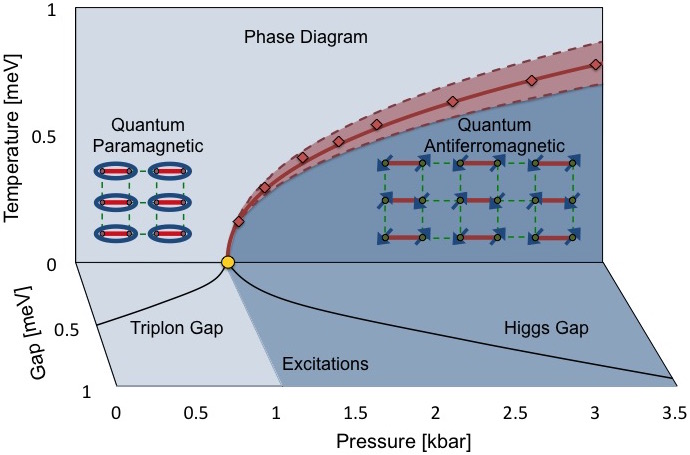}}
\caption{\it The phase/energy diagram of  TlCuCl$_3$ \cite{Merchant}.
The vertical panel shows the pressure-temperature phase diagram, 
the N\'eel temperature curve separates
magnetically ordered and magnetically disordered phases.
The light red band around the N\'eel curve indicates the region of dimensional 
crossover.
The horizontal panel shows both the triplon gap $\Delta_t$
in the paramagnetic phase
and the Higgs magnon gap $\Delta_H$ in the antiferromagnetic phase
versus pressure at zero temperature. 
 }
\label{phase}
\end{figure}
%%%%%%%%%%%%%%%%%%%%%%%%%%%%
The phase diagram of the dimerized 3D quantum antiferromagnet TlCuCl$_3$
is shown in the vertical panel of Fig. \ref{phase}. 
The disordered quantum state consists of an array of spin dimers 
(spin singlets),
and the ordered quantum state has a long range N\'eel order as
illustrated in Fig. \ref{phase}.
The N\'eel temperature curve (red line) separates ordered and disordered phases,
the QCP is indicated by the yellow dot.

\begin{figure}[t!]
 {\includegraphics[width=0.45\textwidth,clip]{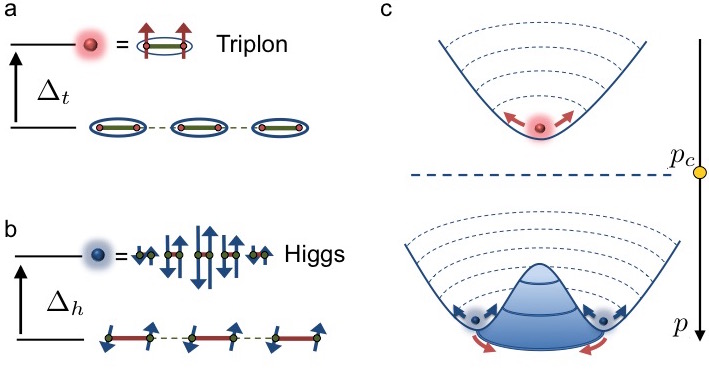}}
\caption{\it  Excitations of a dimerized quantum antiferromagnet.
Panel a illustrates the triple degenerate gapped triplon excitations.
Panel b illustrates the gapped
longitudinal (Higgs) excitation. Panel c illustrates the quantum phase transition; the strength of the interactions in either phase is depicted by the steepness of the `well'. Within the ordered phase, the `Mexican hat' potential has a flat direction which supports the gapless Goldstone excitations (red arrows).  Precisely at the QCP (dashed line), all directions flatten - the Higgs and triplon excitations become gapless and non-interacting \textit{i.e.} asymptotically free.
 }
\label{excit}
\end{figure}

Excitations in the disordered phase, triplons, are gapped.
These are triplet excitations of spin dimers, see Fig. \ref{excit}a. 
There are two different kinds of excitations in the ordered phase,
gapless Goldstone excitations and the gapped longitudinal Higgs excitation,
they are illustrated in Fig. \ref{excit}b \& c.
The horizontal panel in  Fig. \ref{phase} displays excitation gaps versus pressure at zero temperature.
%%%%%%%%%%%%%%%%
%%%%%%%%%%%%%%%%%
Overall the experimental data \cite{Merchant, Ruegg2004, Ruegg2008} provide the following information,
(i) The N\'eel temperature versus pressure, (ii) The magnetic excitation
gap in the disordered phase for various temperatures and pressures,
(iii) The Higgs magnon excitation gap in the 
antiferromagnetic phase for various temperatures and pressures,
(iv) The magnetic excitation width (lifetime) for various temperatures and 
pressures.
In our analysis we do not use fully the width data, the width is used
only to indicate the dimensional crossover region around the N\'eel temperature.
However, we fully use the data from points (i), (ii) \& (iii).
In the analysis we neglect small spin-orbit anisotropy which leads to
a small gap in one of the ``Goldstone'' modes in the antiferromagnetic phase. 

The quantum phase transition (QPT) between ordered and disordered phases
is described by the effective field theory with the following Lagrangian
\cite{Sachdev10,Sachdev11,Kulik,Oitmaa},
\begin{align}   
\label{Lagrangian}
{\cal L}&=\frac{1}{2}\partial_{\mu}{\vec{\varphi}}\partial^{\mu}{\vec{\varphi}}-\frac{1}{2}m^2_0{\vec{\varphi}}^{\ 2}-\frac{1}{4}\alpha_0[\vec{\varphi}^{\ 2}]^{2}.
\end{align}
The vector field $\vec{\varphi}$ describes the staggered magnetisation,
the index $\mu$ enumerates time and three coordinates.
The QPT results from tuning the mass term, $m^2_0$, for which we take the 
linear expansion $m^2_0(p)=\gamma^2(p_c-p)$, where $\gamma^2>0$ is a 
coefficient and $p$ is the applied pressure. Varying the pressure leads 
to two distinct phases; (i) for $p<p_c$ we have $m^2_0>0$, and the classical 
expectation value of the field is zero $\varphi_c^2=0$. This describes the 
magnetically disordered phase, 
the system has a global rotational symmetry, and the excitations are 
gapped and triply degenerate. These excitations are referred to as ``triplons". (ii) For pressures $p>p_c$ we have $m^2_0<0$, 
and the field obtains a non-zero classical expectation value 
$\varphi^2_c=\frac{|m^2_0|}{\alpha_0}$.
This describes the magnetically ordered, antiferromagnetic phase. 
Varying $m^2_0$ from positive to negative spontaneously breaks the O(3) 
symmetry  of the system. In the broken phase  there are 
two gapless transverse (Goldstone \cite{Goldstone}) excitations, and one 
gapped longitudinal (Higgs) excitation. One easily recovers the known 
relation;  Higgs gap/triplon gap$=\sqrt{2}$ \cite{MassRatio}, explicitly $\Delta_t(p)=m_0(p)$ and $\Delta_h(p)=\sqrt{2}|m_0(p)|$.
%%%%%%%%%%%%%%%%%%%
\begin{figure}[Hh]
 {\includegraphics[width=0.48\textwidth,clip]{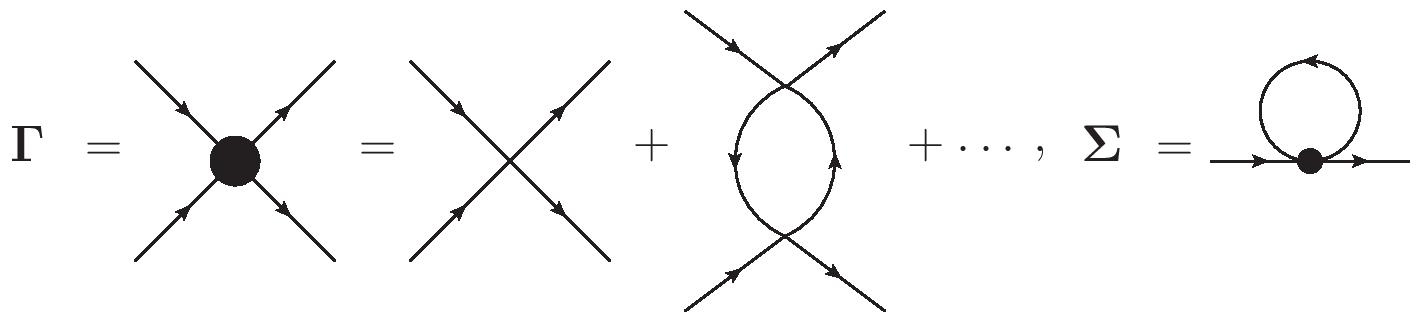}}
\caption{\it Diagrams for the vertex $\Gamma$ and self-energy $\Sigma$.}
\label{figvert}
\end{figure}
%%%%%%%%%%%%%%%%%%%%

The above analysis does not account for quantum or thermal fluctuations.
All fluctuations considered in the present paper originate from
the vertex and self-energy diagrams shown in Fig. \ref{figvert}.
The vertex corrections result in the running coupling constant $\alpha_{\Lambda}$,
see e.g. Ref. \cite{Zinn-Justin} or the Supplemental Material,
\begin{equation}
\label{running}
\alpha_{\Lambda}=\frac{\alpha_0}{1+
\frac{11\alpha_0}{8\pi^2}\ln\left({\Lambda_0}/{\Lambda}\right)} \ .
\end{equation}
Here $\Lambda$ is the energy/momentum scale, and $\Lambda_0$
is the normalization point, $\alpha_{\Lambda_0}=\alpha_0$.
Eq. (\ref{running}) has been obtained within the single loop
renormalization group which implies that $\alpha_0/8\pi \ll 1$.
At the same time the logarithmically enhanced denominator in 
(\ref{running}) can be arbitrarily large.
Note that the normalization point $\Lambda_0$ can be arbitrary, 
generally it is not equal to the ultraviolet cutoff related to the
lattice spacing.
Eq. (\ref{running}) has a pole at 
$\Lambda=\Lambda_L=\Lambda_0e^{8\pi^2/11\alpha_0}$. This is the famous Landau pole
much debated in quantum field theory~\cite{Landau}.
Physics in the vicinity of the Landau pole remains a controversial issue
in spite of 60 years of theoretical studies.
Remarkably quantum magnets can shed light on the problem, we will return to 
this point later.
%%%%%%%%%%%%%%%%%%%
\begin{figure}[Hh]
 {\includegraphics[width=0.33\textwidth,clip]{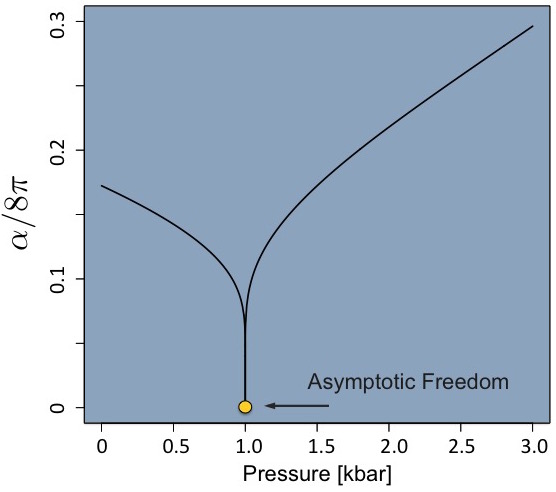}}
\caption{\it Zero temperature running coupling constant versus pressure
in TlCuCl$_3$. The constant vanishes at the QCP (yellow point). 
}    
\label{figg}
\end{figure}
%%%%%%%%%%%%%%%%%%%%
The running coupling constant at zero temperature versus pressure
is plotted in Fig. \ref{figg}. This curve is extracted from available
experimental data.
The coupling constant vanishes at the QCP, indicating
the asymptotic freedom. Fig. \ref{figg} represents one of our central results,
below we explain how we obtain it.
%%%%%%

At zero temperature, equations for the running mass and for the running 
staggered magnetization are well known \cite{Zinn-Justin}, (see also the Supplemental Material)
\begin{align}
\label{t0}
{m}^2(p,\Lambda)&=\gamma^2(p_c-p)\left[\frac{\alpha_{\Lambda}}
{\alpha_0}\right]^{\frac{5}{11}} \\
\label{t1}
\varphi_c^2(p,\Lambda)&=\frac{\gamma^2(p-p_c)}{\alpha_0}
\left[\frac{\alpha_0}{\alpha_{\Lambda}}\right]^{\frac{6}{11}}.
\end{align}
To find actual values of the gap in the disordered phase one has to solve
Eq. (\ref{t0}) at $p < p_c$ with $\Delta_t=\Lambda=m$.
To find the Higgs gap in the ordered phase one has 
to solve Eq. (\ref{t0}) at $p > p_c$ with $\Delta_H=\Lambda=\sqrt{2}|m|$.
The relation $\Delta_H/\Delta_t=\sqrt{2}$ \cite{MassRatio} remains valid
with logarithmic accuracy.

We need to extend the theory to nonzero temperatures.
Our goal is to find excitation spectra, therefore we cannot use the imaginary 
time Matsubara technique. We need to work with real frequencies at a nonzero
temperature. Generally, there is not a regular diagrammatic technique which 
allows to calculate real frequency Green's functions at nonzero temperature.
Fortunately, in the present case such a calculation is possible within
standard techniques.
This possibility comes from the two following observations. (i)
Multi-loop logarithmic corrections are universal, they are not sensitive as to 
whether frequency/energy is real or imaginary.
(ii) The leading in $\alpha$ correction which contains powers of temperature
comes only from the self energy diagram shown in Fig. \ref{figvert}.
Calculation of this diagram does not cause problems since while
it depends on temperature, it is frequency independent.
Still, there is a minor complication related to point (ii).
The complication is due to temperature broadening (quasipraticle lifetime)
because of scattering from the heat bath of magnons. 
Below we explain how we address this issue. To be specific let us 
consider the triplon gap in the disordered phase.
Calculation of the self-energy $\Sigma$, Fig. \ref{figvert}, 
gives the following answer (see the Supplemental Material)
\begin{equation}
\label{dT}
\Delta^2_t(p,T,\Lambda)=\gamma^2(p_c-p)\left[\frac{\alpha_{\Lambda}}
{\alpha_0}\right]^{\frac{5}{11}}+5\alpha_{\Lambda}
\sum_{\bf k}\frac{1/\omega_{\bm k}}{e^{\frac{{\omega}_{\bm k}}{T}}-1}\ .
\end{equation}
At zero temperature the second term in the right hand side is zero 
and Eq. (\ref{dT}) becomes identical to Eq. (\ref{t0}).
The value of the lower logarithmic cutoff in (\ref{dT}) is obvious, 
$\Lambda=\max\{\Delta_t,T\}$. The triplon dispersion is harder.
The naive formula $\omega_{\bm k}=\sqrt{{\bm k}^2+\Delta_t^2}$ is incomplete because at small $k$ and close to the N\'eel temperature where 
$\Delta_t \to 0$ the line width $\Gamma_t$ (temperature broadening)
becomes larger than the gap. Physically the inequality
$\Gamma_t > \Delta_t$ is an indication of the dimensional crossover, 
$4D \to 3D$. Sufficiently close to the N\'eel temperature, critical indices take the 3D 
classical values.
To fix the problem we take 
$\omega_{\bm k}=\sqrt{{\bm k}^2+\Delta_t^2 + \Gamma^2_t}$, this is a standard way
 to describe a damped harmonic oscillator, see \textit{e.g.} Ref. \cite{Merchant}.
Of course the modified dispersion is not sufficient to fully describe the
dimensional crossover, but it is sufficient for the purposes of the present work.
The line broadening we take directly from experiment,
$\Gamma_t=\xi T$, where $\xi \approx 0.15$ \cite{Merchant}.
Solution of Eq. (\ref{dT}) with $\Delta_t=0$ gives the N\'eel temperature
as function of pressure, $T_N(p)$.

One can also approach the N\'eel temperature from the ordered phase, see Supplemental Material.
In this case there are two Goldstone and one Higgs mode.
Equation for the Higgs gap is similar to (\ref{dT}),
\begin{eqnarray}
\label{dTH}
\Delta^2_H(p,T,\Lambda)&&=2\left\{\gamma^2(p-p_c)\left[\frac{\alpha_{\Lambda}}
{\alpha_0}\right]^{\frac{5}{11}}\right.\\
&&\left.-2\alpha_{\Lambda}\sum_{\bf k}\frac{1/k}{e^{\frac{k}{T}}-1}
-3\alpha_{\Lambda}\sum_{\bf k}
\frac{1/{\omega}_{\bm k}}{e^{\frac{{\omega}_{\bm k}}{T}}-1}\right\}.\nonumber
\end{eqnarray}
Again, $\Lambda=\max\{\Delta_H,T\}$,
$\omega_{\bm k}=\sqrt{{\bm k}^2+\Delta_H^2+\Gamma_H^2}$,
and $\Gamma_H=\zeta T$.
The N\'eel temperature determined from the condition $\Delta_t=0$, 
Eq. (\ref{dT}), must be identical to that determined from
$\Delta_H=0$, Eq. (\ref{dTH}). From here we conclude that broadening
of the Higgs mode is larger than that  of the triplon, $\zeta \approx 0.3$
compared to $\xi \approx 0.15$. The larger broadening is consistent with the
data \cite{Merchant}. We note that the critical exponent of the magnetisation in Eq. (\ref{t1}) is identical to that found for the N\'eel temperature by solving Eq. (\ref{dT}) or (\ref{dTH}). This agrees with the latest Quantum Monte Carlo simulations on the three dimensional dimerised antiferromagnet \cite{Qin}. 

%%%%%%%%%%%%%%%%%%%
\begin{figure}[h!]
 {\includegraphics[width=0.33\textwidth,clip]{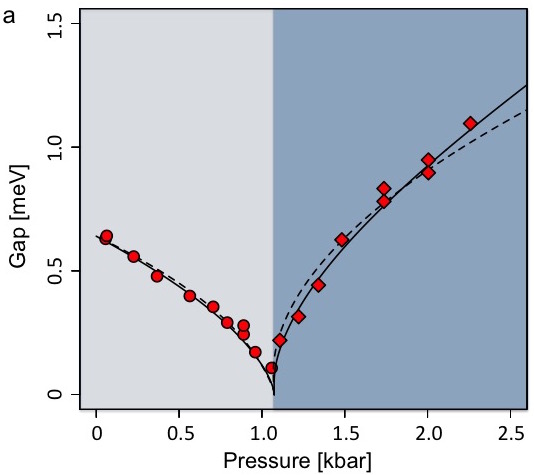}}
 {\includegraphics[width=0.33\textwidth,clip]{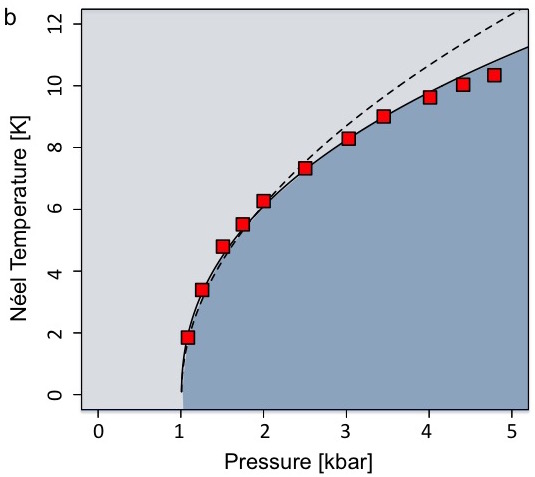}}
\caption{\it Panel a: triplon and Higgs gaps versus presssure
at temperature T=1.85K. Points show experimental data from Ref \cite{Ruegg2008}.
Panel b: N\'eel temperature versus pressure.
Points show experimental data from Ref \cite{Ruegg2004}. In both plots the solid and dashed curves are quantum field theory fits with and without account of the running coupling constant, respectively.
}    
\label{fits}
\end{figure}
%%%%%%%%%%%%%%%%%%%%

Now we are fully armed to perform fits to the experimental data.
As the normalisation point we take $\Lambda_0=1$ meV.
We remind the reader that this choice is arbitrary, one can always use a different
normalisation point with an appropriate rescaling of the coupling
$\alpha_0$.
There are three fitting parameters, the critical pressure $p_c$,
the coefficient $\gamma^2$ in the pressure dependence of the bare mass,
and the coupling constant $\alpha_0$.
Points in panel a of Fig. \ref{fits} show experimental values of the 
triplon and Higgs gaps for various pressures at $T=1.85$ K.
Points in panel b of Fig. \ref{fits} show N\'eel temperatures for various pressures. The data are taken from Refs \cite{Merchant, Ruegg2004, Ruegg2008}.

Solid curves in both panels show fit of the data
with Eq.'s (\ref{running}), (\ref{dT}) and (\ref{dTH}). 
Values of the fitting parameters are 
\begin{equation}
\label{parameters}
p_c=1.01\ \text{kbar}, \  \gamma = 0.68\ \text{meV/kbar$^{1/2}$}, \
\frac{\alpha_0}{8\pi} =0.23 \ .
\end{equation}
It is worth noting that while $T=1.85$K is a pretty low temperature,
the temperature corrections in Eq.'s (\ref{dT}) and (\ref{dTH})
are not negligible.
In the present work we set the triplon speed
equal to unity. If one restores three different speeds along three
different principal directions of the lattice, $c_1$, $c_2$, $c_3$,
then Eq. (\ref{parameters}) is changed to $\alpha_0/(8\pi c_1c_2c_3)=0.23$.
This value is close to the value $\alpha_0/(8\pi c_1c_2c_3)=0.21$ obtained 
in Ref. \cite{Kulik} from the lifetime of the Higgs mode.
The work \cite{Kulik} did not account for the running coupling constant,
however experimentally the major contribution to the lifetime data
comes from Higgs magnons with energy of about 1meV. This energy is taken as
the normalisation point in the present work. 
Coincidence of the energy scales explains the very close agreement between
the accurate result of the present work and the approximate result
of Ref. \cite{Kulik}.

\begin{figure}[Ht]
 {\includegraphics[width=0.33\textwidth,clip]{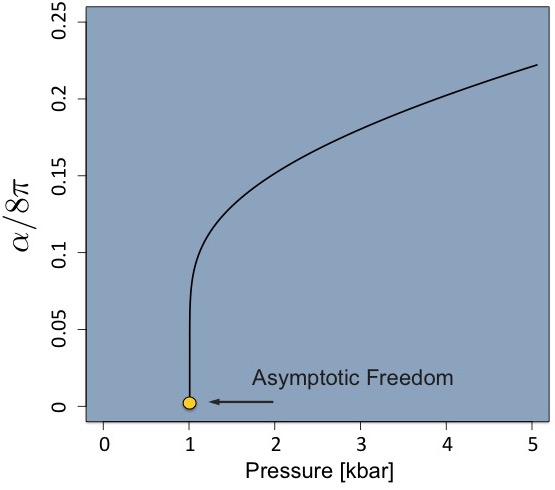}}
\caption{\it Running coupling constant versus pressure
along the N\'eel temperature curve. 
Unlike  Fig. 4 where the temperature is zero, in this case
$T=T_N(p)$. According to Eq.'s (\ref{dT}) and (\ref{dTH}),
the infrared cutoff in Eq. (\ref{running}) for $\alpha_{\Lambda}$ 
is $\Lambda=T_N(p)$. The QCP is again marked by the yellow dot.
}    
\label{at}
\end{figure}

We stress that the coupling constant significantly changes along the 
fitting curves in Fig. \ref{fits}. To illustrate this change we present Fig. \ref{at}, which shows
how the constant runs along the N\'eel temperature curve. 
%%%%%%%%%%%%%%%%%%%
%%%%%%%%%%%%%%%%%%%%
A similar run is shown in  Fig. \ref{figg} where the coupling constant
is plotted versus pressure at zero temperature. In this case 
the infrared cutoff in Eq. (\ref{running}) for $\alpha_{\Lambda}$
is equal to  triplon/Higgs gap at zero temperature.
Position of the Landau pole follows from the known value of the
coupling constant, $\Lambda_L=\Lambda_0e^{8\pi^2/11\alpha_0}\approx 3.5$meV.
This energy is higher than the experimentally studied regime
and is comparable with expected ultraviolet cutoff related
to the dispersion along the third axis, see discussion in Ref. \cite{Kulik}.
Experimental and theoretical studies in this energy range can shed light
on the Landau pole physics and on the expected dimensional crossover in 
TlCuCl$_3$.
Alternatively the Landau pole physics can be addressed in Quantum
Monte Carlo studies of dimerized spin-lattice models~\cite{Qin, And}.

Our central goal is to pin down the running coupling constant which vanishes
at the QCP giving rise to asymptotic freedom. 
To check this statement we have also performed a fit of data with fixed
coupling constant. For this purpose we use the same Eq's. (\ref{dT}) and 
(\ref{dTH})
where we set $\alpha_{\Lambda}=\alpha_0$. The best fitting parameters
are $p_c=1.01$kbar,  $\gamma = 0.64$meV/kbar$^{1/2}$,  $\alpha_0/8\pi =0.16$.
Corresponding fitting curves are shown in Fig. \ref{fits} by dashed lines. 
It is seen from the quality of the fits that the running coupling plays a crucial role in describing the static properties of the system; the analysis clearly demonstrates the running coupling constant.

Asymptotic freedom is a prominent physical phenomenon.
It is the remarkable experimental control of the quantum antiferromagnet 
TlCuCl$_3$ that has allowed the present work to identify for the first time 
the logarithmic decay of the coupling constant.
More generally, with such remarkable experimental control, TlCuCl$_3$ and 
other quantum antiferromagnets provide an ideal playground for 
studies of the Landau pole physics and many other nontrivial
quantum  phenomena.

We thank Bruce Normand, Anders Sandvik and Yaroslav Kharkov for many useful discussions. We also thank Christian R\"uegg for his comments regarding the decay widths. 

\subsection{A. Running Coupling Constant}
\noindent The four point vertex in Fig. (3) is calculated to second order in $\alpha$ (with a Euclidean metric)
\begin{align*}
\Gamma^{(4)}&=\alpha-11\alpha^2\int^{\Lambda_c}_{\Lambda}\frac{d^4k}{(2\pi)^4}\frac{1}{k^4}\\
\tag{A.1}&=\alpha-\frac{11\alpha^2}{8\pi^2}\ln\left(\frac{\Lambda_c}{\Lambda}\right).
\end{align*}
The infrared cut-off, $\Lambda$, is given by the mass gap, or the temperature scale.
We use a Callan-Symanzik equation to find the Beta function
\begin{align*}
\left[\frac{d}{d\ln(\Lambda_c/\Lambda)}+\beta(\alpha)\frac{d}{d\alpha}\right]\Gamma^{(4)}&=0\\
\Rightarrow\beta(\alpha)&=-\frac{11\alpha^2}{8\pi^2}\\
\Rightarrow\frac{d\alpha}{d\ln(\Lambda_0/\Lambda)}&=-\frac{11\alpha^2}{8\pi^2}\\
\label{RunningAlpha}
\tag{A.2}\alpha_{\Lambda}&=\frac{\alpha_0}{1+\frac{11\alpha_0}{8\pi^2}\ln(\Lambda_0/\Lambda)}
\end{align*} 
where $\Lambda_c$ is some momentum cut-off such as the inverse lattice spacing, while $\Lambda_0$ is the `normalization' scale or point. 

\subsection{B. Self-Energy in the Disordered Phase}
\noindent Approaching from the disordered phase, the first perturbative correction to the triplon gap comes from the one loop self energy
\begin{align*}
\label{oneloop}
\tag{B.1}\Sigma(\Delta,T)&=5\alpha_{\Lambda}
\sum_{\bf k}\frac{1}{\omega_{\bm k}}\left[\frac{1}{2}+\frac{1}{e^{\frac{{\omega}_{\bm k}}{T}}-1}\right]\\
&=5\alpha_{\Lambda}\int\frac{d^3k}{(2\pi)^3}\frac{1}{2\omega_{\bm k}}+5\alpha_{\Lambda}\int\frac{d^3k}{(2\pi)^3}\frac{1}{\omega_{\bm k}}\frac{1}{(e^{\frac{\omega_{\bm k}}{T}}-1)}.
\end{align*}
The coupling constant coefficient is the running coupling $\alpha_{\Lambda}$, since the two point corrections are multiplicative with the four point vertices.
With these corrections the triplon gap becomes dependent on both $p$ and $T$
\begin{align*}
\label{masszeroT}
\tag{B.2}\Delta^2(p,T)=m^2_0(p)+\Sigma(\Delta,T).
\end{align*}
The first term in the self energy eq.(\ref{oneloop}) renormalizes the bare mass term $m_0^2$, such that $m_0^2+5\alpha_{\Lambda}\int\frac{d^3k}{(2\pi)^3}\frac{1}{2\omega_{\bm k}}\to m_{\Lambda}^2$ has logarithmic dependence on the energy scale $\Lambda$. After RG, this part results in Eq.(3). The second term, or the `temperature perturbation', only contributes to the logarithmic running via its influence on the infrared cutoff. To make these statements more clear, consider zero temperature such that only the first term contributes. We write the two point function as
\begin{align*}
\Gamma^{(2)}&=m^2+5\alpha_{\Lambda}\int^{\Lambda_c}_{\Lambda}\frac{d^3k}{(2\pi)^3}\frac{1}{2\sqrt{k^2+m^2}}\\
\tag{B.3}&=m^2-\frac{5\alpha_{\Lambda} m^2}{8\pi^2}\ln\left(\frac{\Lambda_c}{\Lambda}\right).
\end{align*}
We use the Callan-Symanzik equation to find the (mass) Beta function
\begin{align*}
\notag0&=\left[\frac{d}{d\ln(\Lambda_c/\Lambda)}+\beta_m(\Lambda)\frac{d}{dm^2}\right]\Gamma^{(2)}\\
\notag\Rightarrow\beta_m(\Lambda)&=-\frac{5\alpha_{\Lambda} m^2}{8\pi^2}\\
\notag\Rightarrow\frac{dm^2}{d\ln(\Lambda_0/\Lambda)}&=-\frac{5\alpha_{\Lambda}m^2}{8\pi^2}\\
\notag&=\left(\frac{-5}{11}\right)\frac{\frac{11}{8\pi^2}\alpha_0}{1+\frac{11\alpha_0}{8\pi^2}\ln(\Lambda_0/\Lambda)}\\
\notag\frac{d\ln(m^2)}{d\ln(\Lambda_0/\Lambda)}&=\left(\frac{-5}{11}\right)\frac{\frac{11}{8\pi^2}\alpha_0}{1+\frac{11\alpha_0}{8\pi^2}\ln(\Lambda_0/\Lambda)}\\
\label{RunningMass}
\tag{B.4}m^2_{\Lambda}&=m_0^2\left(\frac{\alpha_\Lambda}{\alpha_0}\right)^{\frac{5}{11}}
\end{align*} 
Including non-zero temperatures does not change the form of the running coupling nor mass eq.'s (\ref{RunningAlpha}, \ref{RunningMass}), but it does shift the infrared cutoff from $m(p)\to\Lambda=$Max$\{\Delta_t(p,T),T\}$

After accounting for how the coupling terms $m^2$ and $\lambda$ depend on the scale, discussion given below, we find that the gap takes the form
\begin{equation*}
\tag{B.6}\Delta^2_t(p,T,\Lambda)=\gamma^2(p_c-p)\left[\frac{\alpha_{\Lambda}}
{\alpha_0}\right]^{\frac{5}{11}}+5\alpha_{\Lambda}
\sum_{\bf k}\frac{1}{\omega_{\bm k}}\frac{1}{e^{\frac{{\omega}_{\bm k}}{T}}-1}
\end{equation*}

\subsection*{C. Self-Energy in the Ordered Phase}
\noindent The ordered phase is induced by the spontaneous breakdown of the O(3) symmetry when $p>p_c$, as discussed in the introduction. By analogy with the result for the mass gap in the disordered phase, we find that the Higgs mass in the ordered phase takes the form 
\begin{align*}
\tag{C.1}\Delta^2_H(p,T,\Lambda)&=2\left\{\gamma^2(p-p_c)\left[\frac{\alpha_{\Lambda}}
{\alpha_0}\right]^{\frac{5}{11}}\right.\\
&\left.-2\alpha_{\Lambda}\sum_{\bf k}\frac{1/k}{(e^{\frac{k}{T}}-1)}
-3\alpha_{\Lambda}\sum_{\bf k}
\frac{1/{\omega}_{\bm k}}{(e^{\frac{{\omega}_{\bm k}}{T}}-1)}\right\}\notag
\end{align*}
Again, $\Lambda=\max\{\Delta_H,T\}$, and the scale dependence is of course the same as in the disordered phase.
It is a delicate task to calculate the self energy contributions to the Higgs gap, since within the ordered phase our calculations at each order in $\lambda$ must preserve the Goldstone theorem. The Goldstone theorem is a direct result of the remaining O(2) symmetry. We outline the procedure here; In the Lagrangian, the field $\vec{\varphi}=(\vec{\pi},\varphi_c+\sigma)$ is shifted such that the minimum of the potential is $\varphi_c$, and the field oscillations about this shifted minimum are the two Goldstone modes $\vec{\pi}$ and the gapped Higgs mode $\sigma$. 

\noindent We can write an effective potential, ${\cal V}$, from the non-derivative terms of the Lagrangian expanded about the the minimum $\varphi_c$
\begin{align*}
\tag{C.2}{\cal V}=-\frac{1}{2}|m^2|(\vec{\pi},\varphi_c+\sigma)^2 +\frac{1}{4}\lambda\left[(\vec{\pi},\varphi_c+\sigma)^2\right]^2
\end{align*}
\noindent The following two conditions must simultaneously hold true to ensure that $\varphi_c$ is indeed the minimum of the potential, and that to any order in $\lambda$, the perturbations respect the O(2) symmetry and so preserve the Goldstone theorem
\begin{align*}
\label{conditions}
\frac{d{\cal V}}{d\vec{\varphi}}\Big|_{\varphi_c}=0, \ \ \ \ \ \ \text{and} \ \ \ \ \ \ 
\tag{C.3}\frac{d^2{\cal V}}{d\vec{\pi}^2}\Big|_{\varphi_c}=0.
\end{align*}
Since we have already obtained the universal scale dependence of $\alpha_{\Lambda}$ and $m_{\Lambda}$, we do not need to repeat the Callan-Symanzik, RG procedure. We just outline how the thermal perturbations are to be treated. Computing the thermal loops explicitly we obtain the first expression
\begin{align*}
\label{tadpole}
\notag\frac{d{\cal V}}{d\vec{\varphi}}\Big|_{\varphi_c}&=\alpha_{\Lambda}\varphi_c^2-|m_{\Lambda}^2| +2\alpha_{\Lambda}\sum_{\bf k}\frac{1/k}{(e^{\frac{k}{T}}-1)}\\
\tag{C.4}&+3\alpha_{\Lambda}\sum_{\bf k}
\frac{1/{\omega}_{\bm k}}{(e^{\frac{{\omega}_{\bm k}}{T}}-1)}=0\\
\label{phi_corrections}
\tag{C.5}\varphi_c^2&=\frac{|m_{\Lambda}^2|}{\alpha_{\Lambda}} -2\sum_{\bf k}\frac{1/k}{(e^{\frac{k}{T}}-1)}
-3\sum_{\bf k}
\frac{1/{\omega}_{\bm k}}{(e^{\frac{{\omega}_{\bm k}}{T}}-1)}
\end{align*}
where the thermal corrections are split into two separate contributions. This is because one type comes from the one loop self-energy with a Higgs propagator, and the other with a Goldstone propagator. The first summation accounts for loops with massless Goldstone propagators, while the second accounts for loops with massive Higgs propagators, so that $\omega^2_{\bm k}={\bm k}^2+\Delta_H(p,T)^2$. We can now find the Higgs gap using the result from Eq.(C.5). Directly computing the one loop corrections to the Higgs gap, we find
\begin{align*}
\Delta_H^2&=\left\{3\alpha_{\Lambda}\varphi_c^2-|m_{\Lambda}^2|\right.\\
&\left. +2\alpha_{\Lambda}\sum_{\bf k}\frac{1/k}{(e^{\frac{k}{T}}-1)}
+3\alpha_{\Lambda}\sum_{\bf k}
\frac{1/{\omega}_{\bm k}}{(e^{\frac{{\omega}_{\bm k}}{T}}-1)}\right\}\\
\tag{C.6}&=2|m_{\Lambda}|^2-4\alpha_{\Lambda}\sum_{\bf k}\frac{1/k}{(e^{\frac{k}{T}}-1)}
-6\alpha_{\Lambda}\sum_{\bf k}
\frac{1/{\omega}_{\bm k}}{(e^{\frac{{\omega}_{\bm k}}{T}}-1)},
\end{align*}
and we have used Eq.(\ref{phi_corrections}) in passing from the first to second lines. We see that $\Delta_H^2=2\alpha_{\Lambda}\varphi_c^2+O(\alpha^2)$ which is equivalent to the result for the Higgs gap Eq.(6). 

\subsection*{D. N\'eel Temperature}
Approaching from the disordered phase, we calculate the N\'eel temperature by solving Eq.(5) for $\Delta_t(p,T_N)=0$
\begin{align}
\label{TneelTriplon}
\tag{D.1}T_N(p)^2&=\frac{\gamma^2(p-p_c)}{5\alpha_0\sum_{\bf y}
\frac{1/{\omega}_{\bm y}}{(e^{\omega_{\bm y}}-1)}
}\left[\frac{\alpha_0}{\alpha_{\Lambda}}\right]^{\frac{6}{11}},
\end{align}
where $\omega_{\bm y}=\sqrt{{\bm y}^2+(\Gamma/T_N)^2}=\sqrt{{\bm y}^2+\xi^2}$. The fit $\Gamma=\xi T$ was discussed in the main text. 

Similarly, we can approach from the ordered phase and calculate the N\'eel temperature
by solving Eq.(6) for $\Delta_H(p,T_N)=0$ 
\begin{align}
\label{TneelHiggs}
\tag{D.2}T_N(p)^2&=\frac{\gamma^2(p-p_c)}{3\alpha_0\sum_{\bf y}
\frac{1/{\bm y}}{(e^{\tilde{\omega}_{\bm y}}-1)}+2\alpha_0\sum_{\bf y}
\frac{1/{\bm y}}{(e^{\bm y}-1)}
}\left[\frac{\alpha_0}{\alpha_{\Lambda}}\right]^{\frac{6}{11}},
\end{align}
here $\tilde{\omega}_{\bm y}=\sqrt{{\bm y}^2+\zeta^2}$, and the two terms in the denominator are due to the Higgs and Goldstone self-energies. Since the phase transition is of second order, Eq.'s (\ref{TneelTriplon}) and (\ref{TneelHiggs}) are equivalent. Clearly without account of the finite line width $\Gamma$, the equations are identical. As discussed in the main text, we approximate $\Gamma_t=\xi T$ and $\Gamma_H=\zeta T$, from the experimental data. This approximation only becomes important in the vicinity of the phase transition $\Delta_{t/H}< \Gamma_{t/H}$. Equating Eq.'s (\ref{TneelTriplon}) and (\ref{TneelHiggs}), we find how $\xi$ and $\zeta$ are related at the phase transition. Again, we take $\xi\approx 0.15$ and $\zeta\approx 0.3$.


\begin{thebibliography}{99}
\bibitem{Gross}  D. J. Gross and F. Wilczek, Ultraviolet behavior of non-abelian gauge theories, 
\textit{Phys. Rev. Lett.} {\bf  30}, 1343 (1973).
\bibitem{Politzer}
H. D. Politzer. Reliable perturbative results for strong interactions, 
\textit{Phys. Rev. Lett.} {\bf 30}, 1346 (1973).
\bibitem{Landau}
L. D. Landau, A. A. Abrikosov, and I. M. Khalatnikov, 
\textit{Dokl. Akad. Nauk SSSR} {\bf 95,} 497, 773, 1177 (1954).
\bibitem{Zinn-Justin} J. Zinn-Justin.  \textit{Quantum Field Theory and Critical Phenomena} (Oxford University Press, Oxford, 2002).
\bibitem{Merchant} P. Merchant, B. Normand, K. W. Kr\"amer, M. Boehm, D. F. McMorrow and Ch. R\"uegg, \textit{Nature Physics} {\bf 10,} 373-379 (2014).
\bibitem{Ruegg2004} Ch. R\"uegg, A. Furrer, D. Sheptyakov, Th. Str\"assle, K. W. Kr\"amer, H.-U. G\"udel, and L. M\'el\'esi. Pressure-induced quantum phase transition in the spin-liquid TlCuCl$_3$. \textit{Phys. Rev. Lett.} {\bf 93,} 257201 (2004).
\bibitem{Ruegg2008} Ch. R\"uegg, B. Normand, M. Matsumoto, A. Furrer, D. F. McMorrow, K. W. Kr\"amer, H. U. G\"udel, S. N. Gvasaliya, H. Mutka, and M. Boehm. Quantum magnets under pressure: controlling elementary excitations in TlCuCl$_3$. \textit{Phys. Rev. Lett.} {\bf 100,} 205701 (2008).
\bibitem{Sachdev10} S. Sachdev. \textit{Understanding Quantum Phase Transitions}, edited by Lincoln D. Carr (Taylor \& Francis, Boca Raton, 2010).
\bibitem{Sachdev11} S. Sachdev. \textit{Quantum Phase Transitions} (Cambridge Univ. Press, 2011).
\bibitem{Kulik} Y. Kulik and O. P. Sushkov,  Width of the longitudinal magnon in the vicinity of the O(3) quantum critical point. \textit{Phys. Rev. B} {\bf 84}, 134418 (2011).
\bibitem{Oitmaa} J. Oitmaa, Y. Kulik, and O. P. Sushkov. Universal finite-temperature properties of a three-dimensional quantum antiferromagnet in the vicinity of a quantum critical point. \textit{Phys. Rev. B} {\bf 85,} 144431 (2012)
\bibitem{Goldstone} J. Goldstone, A. Salam and S. Weinberg. Broken symmetries. \textit{Phys. Rev.} {\bf 127,} 965-970 (1962).
\bibitem{MassRatio} S. Sachdev. Rapporteur presentation at the 24th Solvay Conference on Physics,\textit{Quantum Theory of Condensed Matter}, Brussels, Oct 11-13, (2008), arXiv:0901.4103
\bibitem{Qin} Y. Q. Qin, B. Normand, A. W. Sandvik and Z. Y. Meng. Multiplicative logarithmic corrections to quantum criticality in three-dimensional dimerized antiferromagnets. arXiv:1506.06073 (2015)
\bibitem{And} S. Jin and A. W. Sandvik.
Universal N\'eel temperature in three-dimensional quantum antiferromagnets.
\textit{Phys. Rev. B} {\bf 85}, 020409(R) (2012).


\end{thebibliography}
\end{document}